\input{aipcheck}

\documentclass[
    ,final           
  ]
  {aipproc}
\usepackage{amsmath}
\usepackage{mathtools}
\usepackage{setspace}
\layoutstyle{6x9}
\usepackage{amsmath}
\usepackage{mathtools}

\begin{document}
\title{Modification of smeared phase transitions by spatial disorder correlations}


\author{David Nozadze, Christopher Svoboda, Fawaz Hrahsheh, and Thomas Vojta}{
  address={Department of Physics, Missouri University of Science and Technology, Rolla, MO 65409, USA}
}

\begin{abstract}
Phase transitions in disordered systems can be smeared if rare spatial regions develop
true static order while the bulk system is  in the disordered phase. Here,
we study the effects of spatial  disorder correlations on such smeared phase transitions.
 The behaviors of  observables are determined within optimal fluctuation theory.
 We show that even short-range correlations  can qualitatively modify
 smeared phase transitions. For positive correlations (like impurity atoms attract
 each other), the order parameter is enhanced, while it is    suppressed for repulsive correlations  (like atoms repel each other). We use computer simulations  to generate various types of disorder correlations, and to  verify our theoretical predictions.
\end{abstract}

\maketitle

\section{Introduction}

Phase transitions  occur  between different macroscopic states of many-particle systems when control parameters such as temperature,
pressure, magnetic field, chemical composition, {\it{etc}}. are varied.
Phase transitions occurring at nonzero temperatures
are called {\it{thermal or classical}} transitions. They are driven by thermal fluctuations.
Another type of   transitions, the so-called {\it{quantum}} phase transitions
occur at absolute zero temperature, when a nonthermal external parameter  is changed \cite{Sachdevbook}.   They are driven by quantum rather than thermal  fluctuations.   The study of observable quantities
at quantum phase transitions is an important  subject of  modern condensed matter theory.

Since materials often feature considerable amounts of quenched disorder, the study
of  phase transitions in the presence of imperfections or randomness has received
much attention. Initially, it was thought that any kind of disorder can destroy
sharp  phase transitions  because (in the presence of disorder) the system
divides itself up into regions which can undergo the transition
at different values of the control parameter. Later, it became clear that
phase transitions are generically  sharp in the presence of weak disorder.
However, under special conditions, rare finite-size spatial regions can indeed undergo the phase transition
independently, leading to a smearing of the transition. This happens, for example,
at quantum phase transitions with overdamped order parameter dynamics \cite{2003_Vojta_PRL, 2006_Vojta_JPhysA, 2010_Vojta_JLTPhys} or
at thermal phase transitions with extended defects  \cite{2003_Vojta_JPA, 2004_Sknepnek_PRB}.

Theories of disordered systems often assume the randomness to be uncorrelated
in space even though real systems contain some amount of disorder correlations.
At critical points, this assumption is justified as long as the correlations are not
too long-ranged. This can be understood, e.g., by
looking at the Harris criterion \cite{1974_Harris_JPSSP} for the stability
of a clean critical point against weak disorder.
Harris divided up
the system  into regions whose linear size is the correlation length  $\xi$.  The  variation of the  local critical temperatures
 from region to region  can be estimated using central limit theorem, yielding  $\delta T_c\sim \xi^{-d/2}$, where $d$ is the space dimension.  Harris observed that the critical  behavior of the clean transition can be unchanged if the fluctuations $\delta T_c$ between the regions  are
 smaller than the global distance $t=T-T_c$ from criticality, {\it{i.e.,}} $\delta T_c<t$.
 Because $\xi\sim t^{-\nu}$, this leads to the inequality $t^{d\nu/2}<t$ for $t\rightarrow 0$, where $\nu$ is the correlation length critical exponent
of the clean system.  Thus, for stability of the clean critical
behavior we must have $d\nu>2$.
This inequality is called the Harris criterion.

Weinrib {\it{et al.,}} \cite{1983_Weinrib_PRB} generalized the  Harris criterion to  long-range correlated disorder. They
considered
power-law  correlations , $C({\bf{x}}-{\bf{y}})=[(T_c({\bf{x}})- T_c)(T_c({\bf{y}})- T_c)]_{\rm{av}}\sim |{\bf{x}}-{\bf{y}}|^{-\alpha}$, where $[...]_{\rm{av}}$ is the disorder average. To estimate fluctuations  between correlation volumes $\xi^d$,  the central limit theorem estimate  has to be replaced by

\begin{equation}\label{AvCor}
 (\delta T_c)^2=\frac{1}{\xi^{2d}}\int_{\xi^d}d{\bf{x}}^d\int_{\xi^d}d{\bf{y}}^dC({\bf{x}}-{\bf{y}})
\sim \xi^{-{\rm{min}}(\alpha ,d)}\,.
\end{equation}
For $\alpha =d$, we find   $(\delta T_c)^2\sim\xi^{-d}  \log \xi  $.
Then, the  stability criterion for clean critical behavior takes the form   $d\nu -2>0$    for $\alpha \geq d$ or
$\alpha \nu-2>0$ for $\alpha <d$.

Thus, for $\alpha \geq d$, the original short-range Harris criterion is recovered.
However, for $\alpha <d$, long-range correlations lead  to a new inequality. Thus, as long as the correlations decay
faster than $x^{-d}$ with distance, correlated and uncorrelated disorder have the same effect on the stability
of the clean critical behavior.

In Ref.~\cite{2012_Svoboda_EPL}, we studied the effects of spatially  correlated disorder on \emph{smeared} phase transitions, and
 we showed that even short-range correlations  qualitatively modify
the behavior of observable quantities. Here, we summarize this theory.
We then report the results of new computer simulations for which we have  generated spatially
correlated atom distributions using   Kawasaki Monte Carlo simulations \cite{1966_Kawasaki_PR}.

\section{Correlated disorder  at smeared phase transitions}
In this section, we summarize the theory of Ref.~\cite{2012_Svoboda_EPL} for the behavior of observables in the tail of
smeared phase transitions.
We consider a material consisting of substances A and B. Material A
is in the magnetic phase
 characterized by a negative distance from criticality $t_{{\rm{A}}}<0$.
Material B is in the paramagnetic phase  with $t_{{\rm{B}}}>0$.
The quantum phase transition from a non-magnetic to a magnetic phase can be tuned by
substituting  magnetic atoms A for nonmagnetic atoms B in the binary alloy A$_{1-x}$B$_{x}$.

Due to statistical fluctuations in the atom distribution, there can be large (rare) spatial regions
 which show  local order because they have a high  local concentration of A atoms even if the
bulk system is in the non-magnetic phase.
If the order parameter fluctuations are overdamped, the quantum dynamics of these regions freezes.
 This leads to inhomogeneous magnetic order and a smearing of the phase transition \cite{2003_Vojta_PRL}.

We now explore the effects of disorder correlations on  smeared phase transitions within
optimal fluctuation theory. Assume that there are attractive   short-range correlations
between like atoms,  with disorder correlation length  $\xi_{\rm{dis}}$. Like atoms
 tend to form clusters of
typical volume  $V_{{\rm{dis}}}\approx 1+a\xi^d_{\rm{dis}}.$  Roughly, the transition point $x^0_c$ where the phase transition would happen
without having any rare regions effects, can be found by setting the average   distance from criticality to zero. This gives
$
x^0_c=- t_{\rm{A}}/(t_{\rm{B}}-t_{\rm{A}}).
$
A rare region of linear size $L_{\rm{RR}}$ can show local static order  if the
local concentration of non-magnetic B atoms is smaller than some critical concentration
 $x_{c}(L_{\rm{RR}})$ which can be estimated from finite size scaling \cite{1983_Barber_Academic},
$
x_c=x^0_c-DL^{-\phi}_{L_{RR}}
$.
Here, $D$ is a non-universal constant, and $\phi$ is the  shift exponent which
takes the value of $1/2$ within mean-field approximation. As the concentration $x_c$
must be  positive, we obtain the condition $L_{\rm{min}}=(D/x^o_c)^{1/\phi}$ for the minimum size
of a locally ordered rare region.

The probability $P(N,N_B)$ for a rare region with  $N=L^{d}_{RR}$ sites to have
$N_B=Nx_{\rm{loc}}$  sites occupied by B atoms is equal to the probability $P_{\rm{clus}}(N/V_{{\rm{dis}}}, N_B/V_{{\rm{dis}}})$
of  finding  $N_B/V_{{\rm{dis}}}$ B-clusters  among $N/V_{{\rm{dis}}}$ clusters. It is given by the binomial distribution
\begin{equation}\label{bin}
P_{\rm{clus}}(N/V_{{\rm{dis}}},x_{\rm{loc}}/V_{{\rm{dis}}})={N/V_{{\rm{dis}}} \choose N_{B}/V_{{\rm{dis}}}}(1-x)^{\frac{N-N_{B}}{V_{{\rm{dis}}}}} x^{\frac{N_{B}}{V_{{\rm{dis}}}}}\,.
\end{equation}
To calculate the total magnetization, we can simply integrate the binomial distribution  (\ref{bin})
over all rare regions showing local order.  In the regime where the concentration $x$ is somewhat larger than $x^0_c$, we find
(up to power-law prefactors)
\begin{equation}
M\sim \exp \left[-\frac{C}{(1+a\xi^d_{\rm{dis}})}\frac{(x-x^0_c)^{2-d/\phi}}{x(1-x)}\right]\label{expo}\,,
\end{equation}
where $C$ is a non-universal constant. In this regime, the magnetization drops exponentially with $x$, and spatial
disorder correlations  modify the prefactor in the exponent.

In the far tail of the smeared transition at $x\rightarrow 1$, the composition dependence of the order parameter
 is given by a non-universal power law $M\sim (1-x)^{\beta} $
with $\beta=(aL^d_{\rm{min}}+a\xi^d_{\rm{dis}})/V_{\rm{dis}}$.
For small disorder correlation length $\xi_{\rm{dis}}\ll L_{\rm{min}}$, earlier results for uncorrelated  disorder \cite{2011_Hrahsheh_PRB}
are recovered,  $\beta \approx L^d_{\rm{min}}$.
In the limit of  large correlation lengths $\xi\gg L_{\rm{min}}$,  all clusters  show local order, and   all magnetic A atoms contribute to the magnetization.
This leads to $\beta=1$.
Thus, in the far tail, disorder correlations modify the exponent of the order parameter.
In both regimes, for attractive correlations as considered here,
 the magnetization $M$ at a given concentration $x$ increases with increasing  correlation length $\xi_{\rm{dis}}$.

\section{Computer simulations}

To illustrate the theory we now show results of
numerical simulations of a toy model. Motivated by the quantum-classical mapping \cite{Sachdevbook} we consider a classical Ising model with infinite-range
interaction in one (timelike) dimension and short-range interaction in  three spacelike dimensions.
The Hamiltonian has the form \cite{2012_Svoboda_EPL}
\begin{equation}\label{ham1}
H = -\frac{1}{L_{\tau}} \sum_{\left< \mathbf{y}, \mathbf{z} \right>, \tau, \tau'} J_0 S_{\mathbf{y}, \tau} S_{\mathbf{z}, \tau'} - \frac{1}{L_{\tau}} \sum_{ \mathbf{y}, \tau, \tau'} J_{\mathbf{y}} S_{\mathbf{y}, \tau} S_{\mathbf{y}, \tau'}\,.
\end{equation}
Here, $\mathbf{y}$ and  $\mathbf{z}$ are space coordinates, $\tau$ is the timelike coordinate.
$L_{\tau}$ is the system size in the time direction. $\left< \mathbf{y}, \mathbf{z} \right>$
denotes pairs of nearest neighbors in space. $J_{\mathbf{y}}$  is a binary-random variable, which takes
values $J_l$ or $J_h$ depending on the type of  atom on the given lattice site $\mathbf{y}$.
The concentration of sites with $J_l$ and $J_h$   are $x$ and $1-x$, respectively.
The values of $J_{\mathbf{y}}$ at different lattice sites are not independent, but they are correlated.

Because the interaction is infinite-ranged  in the timelike direction,  this dimension can be treated
within mean-field theory. This leads to  coupled mean-field equations for the local magnetization
at site ${\mathbf{y}}$
\begin{equation}\label{mf1}
m_{\mathbf{y}} = \tanh{ \beta \left( J_{\mathbf{y}} m_{\mathbf{y}} + \textstyle{\sum}_{\mathbf{z}} J_0 m_{\mathbf{z}} + h \right)}\,,
\end{equation}
 where $h=10^{-10}$ is a small symmetry-breaking field and the sum is over all nearest neighbors of site $\mathbf{y}$.
 $\beta=1/T_{cl}$ is the inverse classical temperature (not related to the physical temperature of the quantum system
 which is encoded in $L_{\tau}$) \cite{Sachdevbook}.
 The smeared phase transition can be tuned by changing the composition $x$ in the temperature range $T_h>T_{cl}>T_l$, where
 $T_h=J_h+6J_0$ and   $T_l=J_l+6J_0$ are the phase transition temperatures for pure systems with all $ J_{\mathbf{y}}=J_h$ and $ J_{\mathbf{y}}=J_l$, respectively.

\begin{figure}
  \includegraphics[width=6.9cm,height=5.4cm]{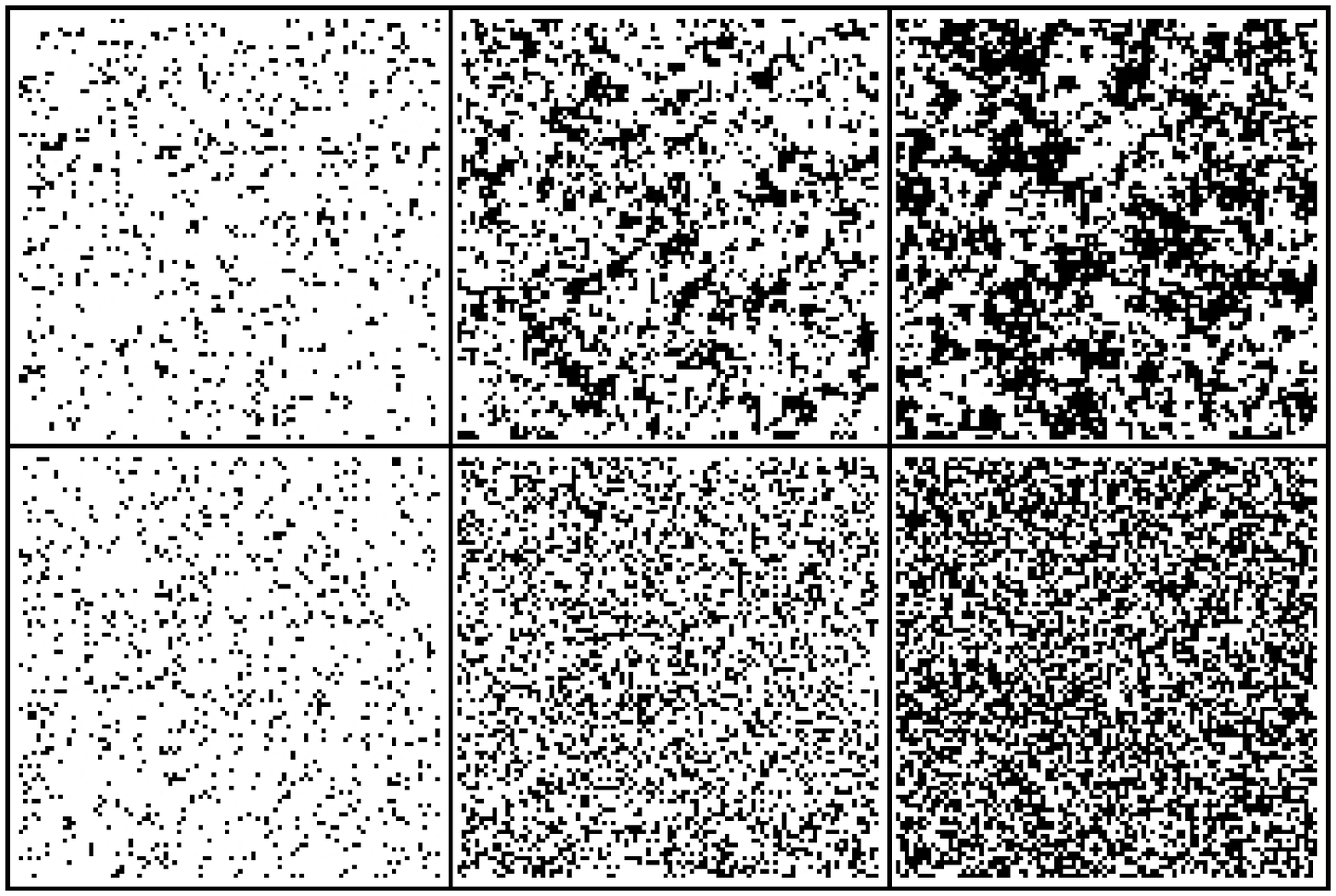}
  \includegraphics[width=7.9cm,height=5.4cm]{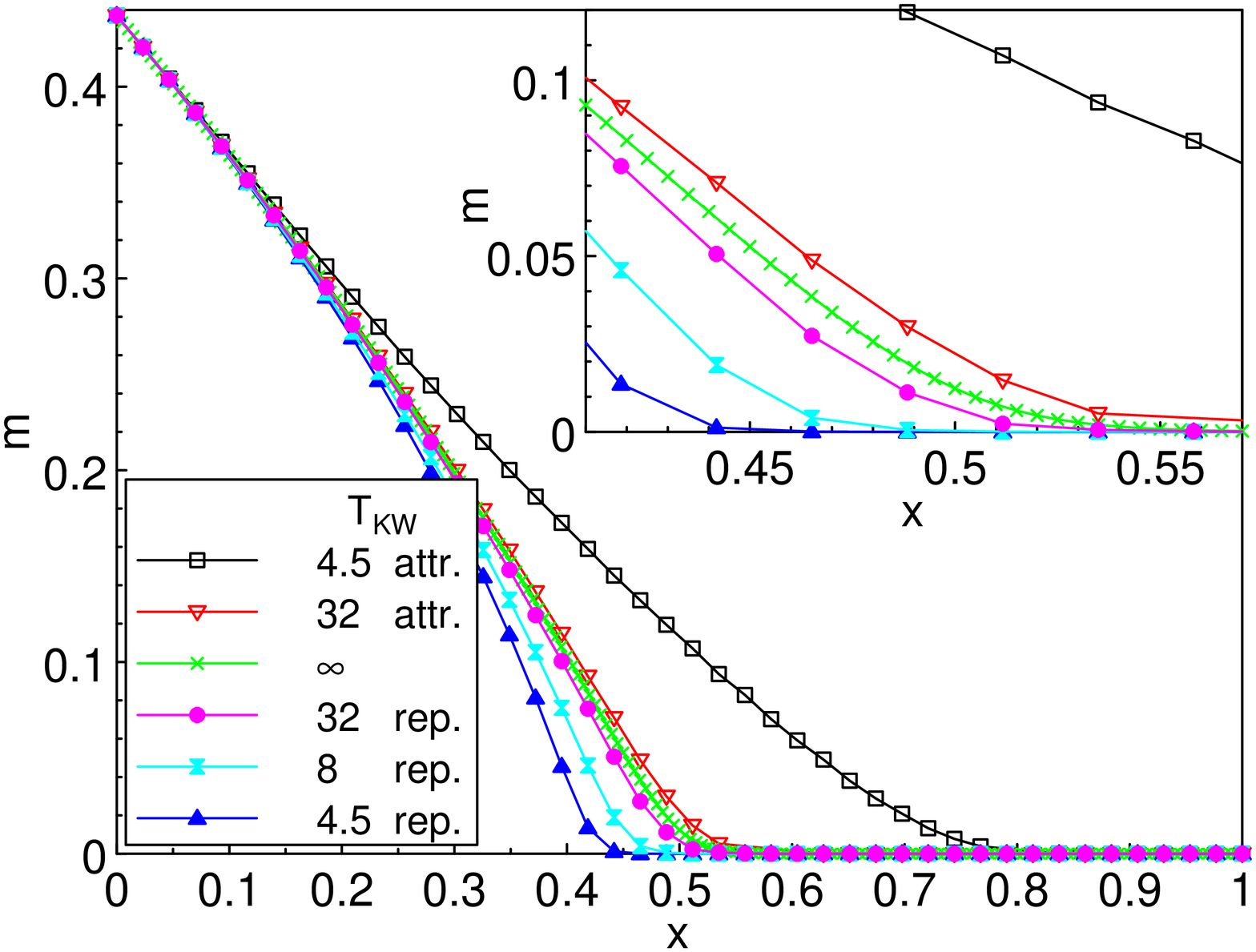}
  \caption{Left: Examples of the atom configuration on a square lattice of $96^2$ sites
generated by Kawasaki Monte Carlo simulations with attractive
interactions $J_{KW}=1$. The Kawasaki temperature $T_{KW}$ is
4.5 in the top row and 32.0 in the bottom row. The concentration $x$
takes values 0.1, 0.3 and 0.5 (left to right). Right: Magnetization M versus composition x for correlated disorder
ranging from strong attractive correlations to strong repulsive (anti-)correlations.
The data stem from one disorder realization of $256^3$ sites for each case, using
$J_h =20$, $J_l =8$, $J_0 =1$, $h=10^{-10}$, and $T_{cl}=24.25$.}\label{snapshots3}
 \end{figure}

 In order to generate correlated binary  random variables we use a more  realistic
 method than the one used in Ref.~\cite{2012_Svoboda_EPL}. We consider a model of $A$ and $B$ atoms with short-range interactions. This model  is equivalent to an Ising model
 \begin{equation}\label{mf2}
H'=-J_{KW}\sum_{\left< \mathbf{y}, \mathbf{z} \right> }n_{\rm{\mathbf{y}}}n_{\rm{\mathbf{z} }}\,,
\end{equation}
 where $n_{\rm{\mathbf{y}}}=\pm 1$ corresponds to $A$ and $B$ atoms and $J_{KW}>0$ for attractive correlations.
 To obtain correlated atom  distributions with a well-defined concentration $x$, we use Monte Carlo simulations by the Kawasaki method \cite{1966_Kawasaki_PR}. We can control the correlations between the atoms by changing the ``Kawasaki"  temperature $T_{KW}$ in these simulations.
 At $T_{KW}=4.5$, atoms are correlated with power-law correlation, while at higher
$T_{KW}$ correlations are  short-ranged (decay exponentially). Correlation length increases from $0$ to $\infty$ with decreasing  $T_{KW}$ from $\infty$ to 4.5. 
 Figure~{\ref{snapshots3}} show
 examples of the atom configurations for two $T_{KW}$  and  several  concentrations $x$.

 To study the smeared phase transition we then solve the  mean-field equations (\ref{mf1}) numerically.
 Figure~{\ref{snapshots3}} show the  magnetization $M$ as a function of concentration $x$ for several disorder correlations (several $T_{KW}$).
 The data show that at a given $x$, the magnetization $M$ increases with decreasing    $T_{KW}$ from
 $\infty$ to 4.5 ({\it{i.e.}} with increasing
 correlation length). Thus, the magnetization is enhanced compared to uncorrelated case.   We also consider repulsive (anti-)correlations.
 To obtain repulsively correlated  atoms, we set $J_{KW}<0$ when generating the atom distribution.   The solution of
 (\ref{mf1}) shown in Figure~{\ref{snapshots3}}  demonstrates that the magnetization is suppressed compared to the uncorrelated case.

\section{Conclusions}

We have studied the effects of spatial correlations between impurity atoms on
 smeared phase transitions.  We have shown that even short-range correlations
between atoms  have dramatic effects and qualitatively modify the behavior of observables.
Attractive disorder correlations  enhance the magnetization in the tail of the smeared  transition, while repulsive correlations suppress it.
This is in contrast to conventional critical behavior, at which  correlations that decay sufficiently fast lead to the same critical behavior
as uncorrelated disorder. We have verified our theoretical predictions by performing computer simulations. To this end, we have generated both attractively and
repulsively  correlated disorder by means of  Kawasaki Monte Carlo simulations. We have found that the resulting data qualitatively agree with our theory.
We emphasize that our results do apply not only to the quantum phase transitions considered here, but  to all disorder induced smeared phase transitions.

\begin{theacknowledgments}
This work has been supported by the NSF under
Grant Nos. DMR-0906566 and DMR-1205803.
\end{theacknowledgments}

\bibliographystyle{aipproc}   


\end{document}